\documentclass[aps,twocolumn,10pt,showpacs,groupedaddress,superscriptaddress,floatfix,notitlepage]{revtex4-2}
\usepackage{amsmath,amsfonts,amssymb,graphics,graphicx,epsfig,color,times, braket}
\usepackage[utf8x]{inputenc}
\usepackage{color}
\usepackage{bbm, dsfont}
\usepackage{subfigure}
\usepackage{mathrsfs}
\usepackage{verbatim}
\usepackage[colorlinks=true, allcolors=blue]{hyperref}
\usepackage{graphicx}
\usepackage{dcolumn}
\usepackage{bm}
\usepackage{physics,amsmath,amssymb,enumitem,hyperref} 
\usepackage{float}

\begin{document}

\preprint{APS/123-QED}

\title{Excitation transfer and many-body dark states in waveguide quantum electrodynamics}

\author{Wei Chen}
\email{maxchen1130@gmail.com}
\affiliation{Department of Physics and Center for Theoretical Physics, National Taiwan University, Taipei, Taiwan}
\affiliation{Institute of Atomic and Molecular Sciences, Academia Sinica, Taipei 10617, Taiwan}
\affiliation{Center for Quantum Science and Engineering, National Taiwan University, Taipei 10617, Taiwan}

\author{Guin-Dar Lin}
\email{guindar.lin@gmail.com}
\affiliation{Department of Physics and Center for Theoretical Physics, National Taiwan University, Taipei, Taiwan}
\affiliation{Center for Quantum Science and Engineering, National Taiwan University, Taipei 10617, Taiwan}
\affiliation{Trapped-Ion Quantum Computing Laboratory, Hon Hai Research Institute, Taipei 11492, Taiwan}
\affiliation{Physics Division, National Center for Theoretical Sciences, Taipei 10617, Taiwan}

\author{H. H. Jen}
\email{sappyjen@gmail.com}
\affiliation{Institute of Atomic and Molecular Sciences, Academia Sinica, Taipei 10617, Taiwan}
\affiliation{Physics Division, National Center for Theoretical Sciences, Taipei 10617, Taiwan}




\date{\today}

\begin{abstract}
In one-dimensional waveguide quantum electrodynamics systems, quantum emitters interact through infinite-range, dispersive, and dissipative dipole-dipole interactions mediated by guided photonic modes. These long-range periodic interactions give rise to rich many-body physics absent in free space. In this work, we construct a set of symmetrized multi-excitation dark states and derive analytic expressions for their time-evolution projections. This framework captures the essential dynamics of excitation transport and storage while significantly reducing computational complexity compared to full master equation simulations. Our analysis reveals a fundamental bound on energy redistribution governed by the structure of dark states and collective dissipation, and discovers that optimal excitation transfer between emitter ensembles converges toward an initial pumped fraction of $N_\text{p}/N \approx 0.55$ for large system sizes. We further examine the robustness of this mechanism under realistic imperfections, including positional disorder, nonradiative decay, and dephasing. These results highlight the role of many-body dark states in enabling efficient and controllable energy transfer, offering new insights into dissipative many-body dynamics in integrated quantum platforms.
\end{abstract}


\maketitle


\section{Introduction}
A central concept in waveguide quantum electrodynamics (WQED) \cite{sheremet2023waveguide, lodahl2017chiral, arcari2014near, le2021experimental, douglas2015quantum, chang2018colloquium, mahmoodian2020dynamics, iversen2022self, gonzalez2024light, le2005nanofiber, kien2008cooperative, van2013photon, goban2015superradiance, pichler2015quantum, shahmoon2016highly, ruostekoski2016emergence, ruostekoski2017arrays} is the formation of collective states—especially dark states—which emerge from coherent superpositions of emitter excitations and can suppress radiative decay via destructive interference. These states enable robust storage and controlled transfer of quantum excitations \cite{holzinger2022control, asenjo2017exponential, plankensteiner2015selective, moreno2019subradiance, masson2020atomic, ferioli2021storage, needham2019subradiance, zanner2022coherent}, making them essential for applications such as quantum memory \cite{lvovsky2009optical, zhao2009long, zhao2009millisecond}, quantum networks \cite{dinc2020multidimensional, almanakly2022towards, stannigel2012driven, ramos2014quantum}, and quantum batteries \cite{quach2022superabsorption, ferraro2018high, quach2020using}. Understanding the dynamics and structure of such collective states is thus crucial, as it opens pathways toward scalable quantum information processing and efficient light–matter interfaces in integrated quantum platforms.

Building on Dicke’s seminal work on cooperative decay \cite{dicke1954coherence, gross1982superradiance}, early studies revealed how subradiant states emerge from interference among emitters in reduced-dimensional reservoirs \cite{albrecht2019subradiant, bettles2016cooperative, ballantine2021quantum}. Initially explored in the single-excitation regime, recent work has extended these ideas to many-body dark states that can store multiple excitations while remaining protected from dissipation \cite{asenjo2017exponential, holzinger2022control, poshakinskiy2021dimerization, masson2020atomic, fasser2024subradiance}. These states are often engineered by placing emitters at specific intervals relative to the guided-mode wavelength, forming degenerate subradiant manifolds \cite{holzinger2022control, fasser2024subradiance}. Parallel advances in superradiance and superabsorption have demonstrated how collective emission and absorption processes can be harnessed and stabilized through dark-state dynamics \cite{higgins2014superabsorption, yang2021realization, quach2022superabsorption}. Experimentally, subradiant and superradiant effects have been observed in cold atoms coupled to nanophotonic waveguides \cite{okaba2019superradiance, corzo2019waveguide} and superconducting qubits in microwave circuits \cite{zanner2022coherent, mirhosseini2019cavity, astafiev2010resonance, honigl2020two}, underscoring the central role of interference-driven collective states in WQED and motivate further exploration of many-body effects in strongly coupled quantum systems.

This manuscript addresses the theoretical challenges of modeling excitation dynamics in WQED systems, where nonlocal and collective emitter-emitter interactions lead to complex non-Hermitian dynamics and an exponentially growing Hilbert space. These features make full master equation simulations computationally prohibitive, particularly for long-time evolution or steady-state analysis. Focusing on a highly symmetric configuration, we construct a set of symmetrized multi-excitation dark states and derive analytic expressions for their time-evolution projections. This framework captures the essential physics of excitation transport and localization while significantly reducing computational cost. Although this method leverages strong permutation symmetry and is not directly generalizable to arbitrary WQED geometries, it provides a powerful and exact solution for this important class of analytically tractable systems. We apply this approach to a model of two coupled emitter ensembles in a one-dimensional (1D) chiral waveguide with infinite-range dipole-dipole interactions (DDI) \cite{solano2017super}, analyze the excitation transfer between ensembles, and identify the optimal pumping configuration for maximal energy redistribution. Finally, we examine the robustness of this mechanism under realistic imperfections, including positional disorder, nonradiative decay, and dephasing.

\section{Theoretical model}
\begin{figure}[t]
    \centering
    \includegraphics[width=1\linewidth]{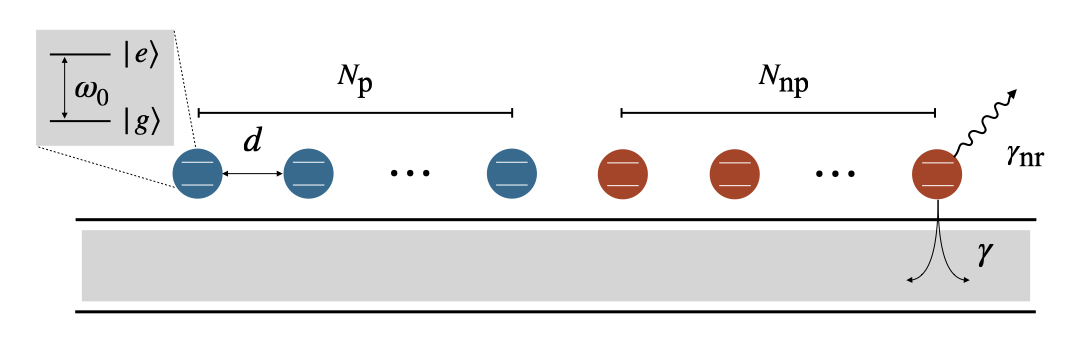}
    \caption{\textbf{Schematic representation of the system setup}. Two 1D emitter arrays, consisting of $N_{\text{p}}$ and $N_{\text{np}}$ emitters, interact via an infinite-range DDI mediated by a 1D waveguide mode. Each emitter is a two-level system with ground state $\ket{g}$, excited state $\ket{e}$, and transition frequency $\omega_0$. The emitters are evenly spaced with a separation of $d$. Without loss of generality, we consider the emitters in the left ensemble (of size $N_{\text{p}}$) are initially pumped to the excited state $\ket{e}$, while those in the right ensemble (of size $N_{\text{np}}$) remain in the ground state $\ket{g}$. In addition to the collective emission into the waveguide (rate $\gamma$), each emitter undergoes nonradiative decay at a rate $\gamma_{\text{nr}}$.}
    \label{system}
\end{figure}

We consider a 1D WQED system consisting of two spatially separated emitter arrays, containing $N_\text{p}$ and $N_\text{np}$ two-level quantum emitters, respectively. These arrays are coupled via infinite-range DDI mediated by waveguide modes, as illustrated in Fig.~\ref{system}. The emitters are positioned along the waveguide with uniform spacing $d$, such that the arrays form a single 1D chain of $N = N_\text{p} + N_\text{np}$ sites labeled by positions $x_i$. The last emitter in the first array is placed and ordered before the first emitter of the second array.

We consider two-level quantum emitters with ground state $\ket{g}$ and excited state $\ket{e}$, and transition frequency $\omega_0$, which is smaller than the cutoff frequency. The system dynamics is governed by the following Markovian master equation for the system’s density matrix $\hat{\rho}$ (with $\hbar = 1$) \cite{lalumiere2013input, chang2012cavity}:
\begin{equation}
    \dot{\hat{\rho}} = -i \left( \hat{\mathcal{H}}_\text{eff} \hat{\rho} - \hat{\rho} \hat{\mathcal{H}}_\text{eff}^\dagger \right) + \sum_{m,n=1}^N \gamma_{m,n} \hat{\sigma}_m \hat{\rho} \hat{\sigma}_n^\dagger,
    \label{master_eq}
\end{equation}
where $\hat{\sigma}_n = \ket{g}_n\bra{e}$ is the lowering operator for the $n$-th emitter, and the non-Hermitian effective Hamiltonian is given by
\begin{equation}
    \hat{\mathcal{H}}_\text{eff} = \sum_{m,n=1}^{N} \left( J_{m,n} - i\frac{\gamma_{m,n}}{2} \right) \hat{\sigma}_m^\dagger \hat{\sigma}_n.
    \label{effective_H}
\end{equation}
Here, $J_{m,n} = (\gamma/2)\sin(k_0 x_{m,n})$ and $\gamma_{m,n} = \gamma \cos(k_0 x_{m,n})$ represent the coherent and dissipative dipole-dipole interactions between emitters $m$ and $n$, respectively, with $x_{m,n} = |x_m - x_n|$, and $k_0 = 2\pi / \lambda_0$ the wavevector with the transition wavelength $\lambda_0$. All emitters are assumed to have the same radiative decay rate $\gamma$ into the waveguide.

We focus on the so-called \emph{mirror configuration}, where the inter-emitter spacing $d = n\lambda_0$ is an integer multiple of the resonant wavelength \cite{holzinger2022control, moreno2019subradiance}. In this regime, the coherent coupling vanishes, $J_{m,n} = 0$, and the dissipative coupling becomes uniform, $\gamma_{m,n} = \gamma$, leading to collective decay behavior. If the initial state is symmetric within each ensemble, the system remains in the symmetric subspace of the Hilbert space throughout its evolution. This allows a substantial reduction in the computational complexity, from the full $2^N$-dimensional Hilbert space to a smaller $(N_\text{p} + 1) \times (N_\text{np} + 1)$ subspace for large system size. Note that, under the mirror configuration, the emitters in different ensembles are actually distinguished by their initial state, instead of their position.

We define collective lowering operators for the two ensembles as $\mathcal{S}_\text{p} = \sum_{n=1}^{N_\text{p}} \hat{\sigma}_n$ and $\mathcal{S}_\text{np} = \sum_{n=N_\text{p}+1}^{N} \hat{\sigma}_n$, which allow the effective Hamiltonian under the mirror configuration to be expressed as
\begin{equation}
    \hat{\mathcal{H}}_\text{eff} = -\frac{i\gamma}{2} \left( 
        \mathcal{S}_\text{p}^\dagger \mathcal{S}_\text{p} 
        + \mathcal{S}_\text{p}^\dagger \mathcal{S}_\text{np}
        + \mathcal{S}_\text{np}^\dagger \mathcal{S}_\text{p}
        + \mathcal{S}_\text{np}^\dagger \mathcal{S}_\text{np}
    \right).
\end{equation}
Note that in this configuration, the Hamiltonian becomes anti-Hermitian, $\hat{\mathcal{H}}_\text{eff} = -\hat{\mathcal{H}}_\text{eff}^\dagger$, reflecting the absence of coherent interactions. For the initial state, we consider all emitters in the first ensemble to be fully excited, while those in the second ensemble are in the ground state, i.e., $\ket{\Psi(0)} = \ket{e}^{\otimes N_\text{p}} \otimes \ket{g}^{\otimes N_\text{np}}$. The system is then left to evolve under the waveguide-mediated collective dissipation until a steady state is reached.

\section{$M$-excitation dark states}
In principle, we may construct the dark states of the system by finding the zero-eigenstates of the effective Hamiltonian in Eq. (\ref{effective_H}) with eigenvalue equal to 0 \cite{holzinger2022control}, e.g., $\hat{H}_\text{eff} \ket{\Psi_\text{D}} = 0$. The simplest form of such zero-eigenstates with $M$ excitations is the product of $M$ pairs of the antisymmetric states (dimers), $\ket{\Psi_\text{D}} \propto (\sigma_i^\dagger - \sigma_j^\dagger)...(\sigma_m^\dagger - \sigma_n^\dagger)\ket{G}$, where $\ket{G} = \ket{g}^{\otimes N}$ \cite{poshakinskiy2021dimerization, holzinger2022control}. Since the emitters in the same ensemble are symmetric, all possible combinations of emitters forming dark states in this form are degenerate. Thus we can obtain the form of the general form of the permutation-symmetric dark state within the $M$-excitation manifold of Hilbert space by symmetrizing those identical emitters (up to normalization):
\begin{align}
\ket{\Psi_D^{(N_p, M)}} \propto   &\sum_{k=0}^{M} (-1)^k
\begin{pmatrix}
N_p - M + k \\
k
\end{pmatrix} \notag \\
\times &
\begin{pmatrix}
N - N_p - k \\
M-k
\end{pmatrix}
\left( \mathcal{S}_\text{p}^\dagger \right)^{M-k} \left( \mathcal{S}_\text{np}^\dagger \right)^k
\ket{G}. \label{DS_expression}
\end{align}
The binomial coefficient in each term of the summation represents the number of possible ways to distribute $M$ excitations between the two ensembles, with $M-k$ excitations in the pumped ensemble and $k$ in the un-pumped one. Eq.~(\ref{DS_expression}) can be interpreted as a combinatorial result that counts the possible configurations of selecting one emitter from each ensemble to form antisymmetric pairs. Further details of the derivation are provided in the Appendix. Since the two ensembles are distinguished by their initial conditions (with the emitters in the pumped ensemble initially fully inverted), the maximum number of excitations that can participate in dark-state formation is constrained by $M \le \min\{N_\text{p}, N_\text{np}\}$—the maximum number of possible dimers in the system. This constraint arises from the initial state preparation rather than from the structure of the Hamiltonian. Note that Eq. (\ref{DS_expression}) reduces to Eq.(S.5) in the Supplemental Material of Ref.~\cite{holzinger2022control} when $M = N_\text{p}$.

 \begin{figure}[t]
    \centering
    \includegraphics[width=1\linewidth]{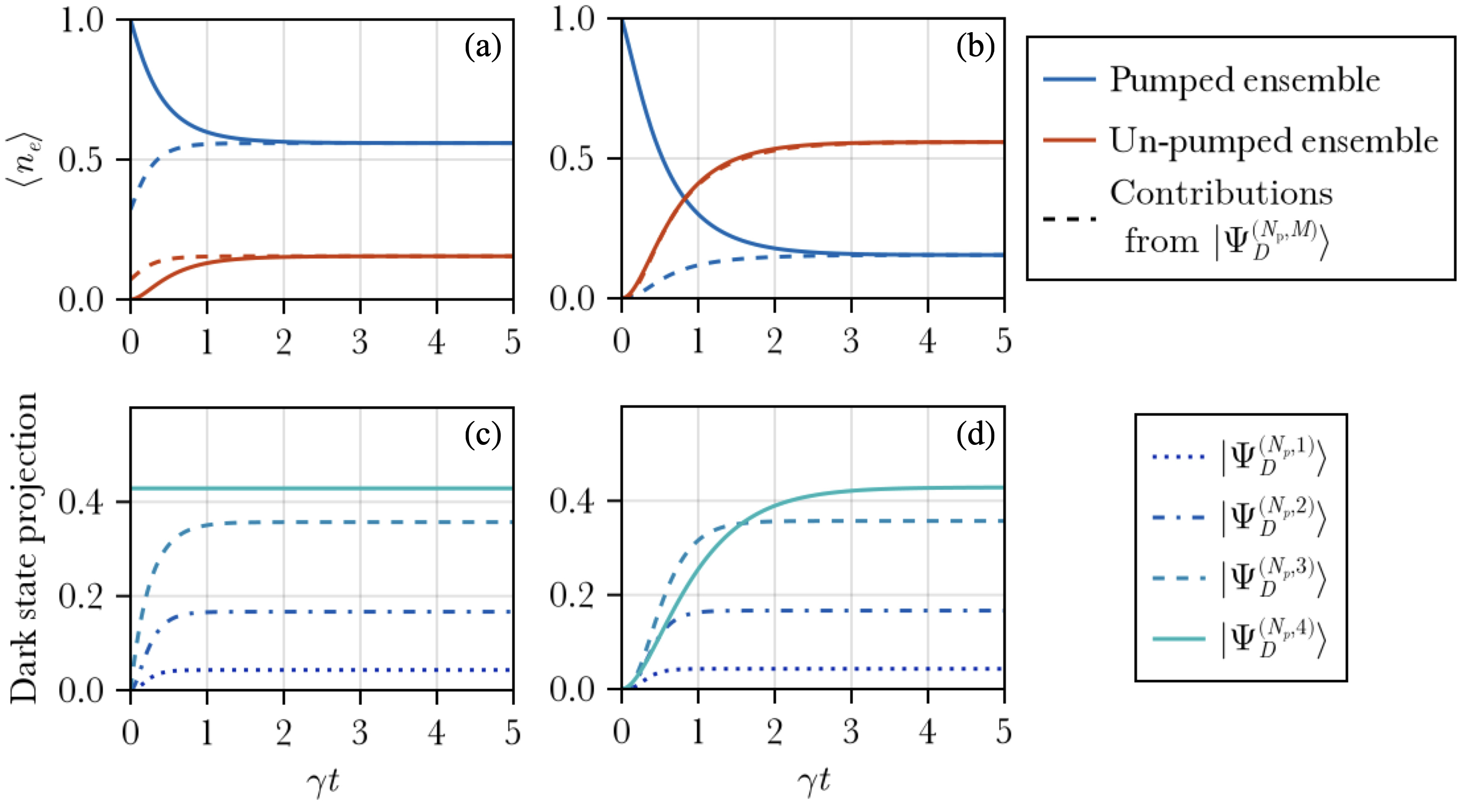}
    \caption{\textbf{Full master equation simulation of the system.} The time evolution of the mean excitation number in each emitter ensemble (red for the pumped ensemble and blue for the un-pumped ensemble), along with the projection of the density matrix onto $M$-excitation dark states. The dashed lines in (a) and (b) represent the expectation values contributed from only the dark-state manifold. In (c) and (d), we separate the projection onto the dark states with different excitation number $M$. In all panels, the total number of emitters is $N = 10$. However, the pumped ensemble size is $N_\text{p} = 4$ in (a) and (c), while it is $N_\text{p} = 6$ in (b) and (d).}
    \label{time_evolution}
\end{figure}

The results of the full master equation simulation of the master equation for $(N, N_\text{p}) = (10, 4)$ and $(N, N_\text{p}) = (10, 6)$, shown in Fig.~\ref{time_evolution}(a, c) and (b,d) respectively, illustrate the time evolution of the mean excitation number in each emitter ensemble (solid lines in panels (a, b)) and the corresponding projections onto the $M$-excitation dark states (lines in panels (c, d)). The dashed lines in Fig.~\ref{time_evolution}(a, b) represent the contributions from only the dark-state manifold to the mean excitation number, which converge to the solid lines in the steady state. As the system evolves, the excitation dynamics of the pumped and un-pumped ensembles exhibit distinct behaviors, with the steady-state values depending on the initial sizes of the two ensembles.

In Fig.~\ref{time_evolution}(c), the initial state has a nonzero projection onto the $(M=N_\text{p})$-excitation dark state. According to the definition of dark states, population remains confined within the dark-state manifold; thus, projections onto lower-excitation dark states increase as the system decays into lower-excitation sectors. A similar trend is observed in Fig.~\ref{time_evolution}(d), but in this case, the initial state has no overlap with the highest excitation ($M = N_\text{np}$) dark state.

Notably, Figs.~\ref{time_evolution}(c) and (d) reveal a symmetry in the dark-state projections for complementary pairs of simulations satisfying $(N_{\text{p},1}, N_{\text{np},1}) = (N_{\text{np},2}, N_{\text{p},2})$, where the steady-state projections remain unchanged. Consequently, the steady-state mean excitation numbers in each ensemble, shown in Fig.~\ref{time_evolution}(a) and (b), are exchanged between these complementary configurations, reflecting the swapped ensemble sizes. These results highlight the interplay between ensemble partitioning and dark-state formation and provide insight into the mechanism of excitation transfer between emitter ensembles.

\section{Analytic solution for dark state projections}To investigate excitation transfer without performing resource-consuming time evolution simulations, we obtained the analytic solution for projection onto each $M$-excitation dark state, $\langle \rho_\text{D}^{M} \rangle$, where $\hat{\rho}_\text{D}^M = |\Psi_\text{D}^{(N_\text{p}, M)}\rangle \langle \Psi_\text{D}^{(N_\text{p}, M)}|$. The equation of motion of the dark-state projection is given by:
\begin{align}
    \frac{d\langle \hat{\rho}^{M}_\text{D} \rangle}{dt} 
    & = \text{Tr} \left[ -i \hat{\rho}_\text{D}^M \left( \hat{\mathcal{H}}_\text{eff} \hat{\rho} -  \hat{\rho} \hat{\mathcal{H}}_\text{eff}^\dagger \right) + \gamma \hat{\rho}_\text{D}^M \mathcal{S} \hat{\rho}\mathcal{S}^\dagger \right], \notag\\
    &= \gamma \langle \mathcal{S}^\dagger \hat{\rho}_\text{D}^{M} \mathcal{S} \rangle, \label{EoM_DS projection}
\end{align}
where $\mathcal{S} \equiv \mathcal{S}_\text{p} + \mathcal{S}_\text{np}$ is the total spin operator. Here, the first term in the right-hand-side of the first line vanishes since $\hat{\rho}_\text{D}^M$ is the density matrix of the zero-eiegenstate of $\hat{\mathcal{H}}_\text{eff}$. From Eq.~(\ref{EoM_DS projection}) we can tell that as the system evolves, the $M$-excitation projection would increase at a rate of $\gamma \langle \mathcal{S}^\dagger \hat{\rho}_\text{D}^{M} \mathcal{S} \rangle$, and depends on the projection onto the state proportional to $\mathcal{S}^\dagger |\Psi_\text{D}^{(N_\text{p}, M)}\rangle$. By defining $\hat{\rho}_\text{D}^{M, k} \equiv (\mathcal{S}^\dagger)^k \hat{\rho}_\text{D}^M(\mathcal{S})^k$, we come to a hierarchy of cascaded equations of motion all the way up to $k = N_\text{p} - M$:
\begin{equation}
    \frac{d\langle \hat{\rho}^{M, k}_\text{D} \rangle}{dt} = -\gamma C_k \left< \hat{\rho}_\text{D}^{M, k} \right> + \gamma \left< \hat{\rho}_\text{D}^{M, k+1} \right>,
    \label{system of ODEs}
\end{equation}
where $C_k = k (C - k)$ and $C = N-2M+1$. Equation (\ref{system of ODEs}) can be obtained by using the commutation relation between the collective operators, $[\mathcal{S}^\dagger, \mathcal{S} ] = 2\mathcal{S}_z$ \cite{gross1982superradiance} and the anti-Hermitian property of the effective Hamiltonian in Eq. (\ref{effective_H}). The integer $k$ ranges from $0$ to $N_\text{p}-M$ since the projection onto $\langle \hat{\rho}_\text{D}^{M, k} \rangle$ for some $k \ge N_\text{p}-M$ is zero all the time, and the recursion stops at $\langle \hat{\rho}_\text{D}^{M, k=0} \rangle = \left< \hat{\rho}_\text{D}^{M} \right>$.

The solution of Eq. (\ref{system of ODEs}) for $N_\text{p} \le N/2$ is given by:
\begin{align}
    \langle \hat{\rho}_\text{D}^{M} \rangle = &\left< \hat{\rho}_\text{D}^{M, \ell=N_\text{p}-M} \right>_0 \notag \\
    & \times \sum_{m = 0}^{\ell} (-1)^\ell \left( \prod_{\substack{n=0 \\ n \ne m}}^{\ell} \frac{1}{C_m - C_n}  \right) e^{-\gamma C_m t}. \label{analytic_sol}
\end{align}
The coefficients $C_m \ne C_n$ for $N_\text{p} \le N/2$. This solution allows us to calculate the excitation distribution among all emitter sites with given system parameters such as $N, N_\text{p}, M $ and the initial expectation value of $\langle \hat{\rho}_\text{D}^{M, \ell=N_\text{p} - M} \rangle_0$. From Eq. (\ref{analytic_sol}), we can derive the steady-state solution by taking the limit of $t \rightarrow \infty$. All the terms except for $m=0$ vanish due to the exponential function in Eq.~(\ref{analytic_sol}):
\begin{equation}
\begin{aligned}
    \langle \hat{\rho}_\text{D}^{M} \rangle_\text{ss} = &\left< \hat{\rho}_\text{D}^{M, \ell=N_\text{p}-M} \right>_0 (-1)^\ell \left( \prod_{\substack{n=1}}^{\ell} \frac{1}{C_0 - C_n}  \right).
\label{ss_analytic_sol}
\end{aligned}
\end{equation}
Due to the symmetry of steady-state dark-state projections for configuration of $N_\text{p} \rightarrow N_\text{np}$ with a fixed $N$, we can obtain the analytical steady-state projection for $N_\text{p} > N/2$ as well.

\begin{figure}[b]
    \centering
    \includegraphics[width=1\linewidth]{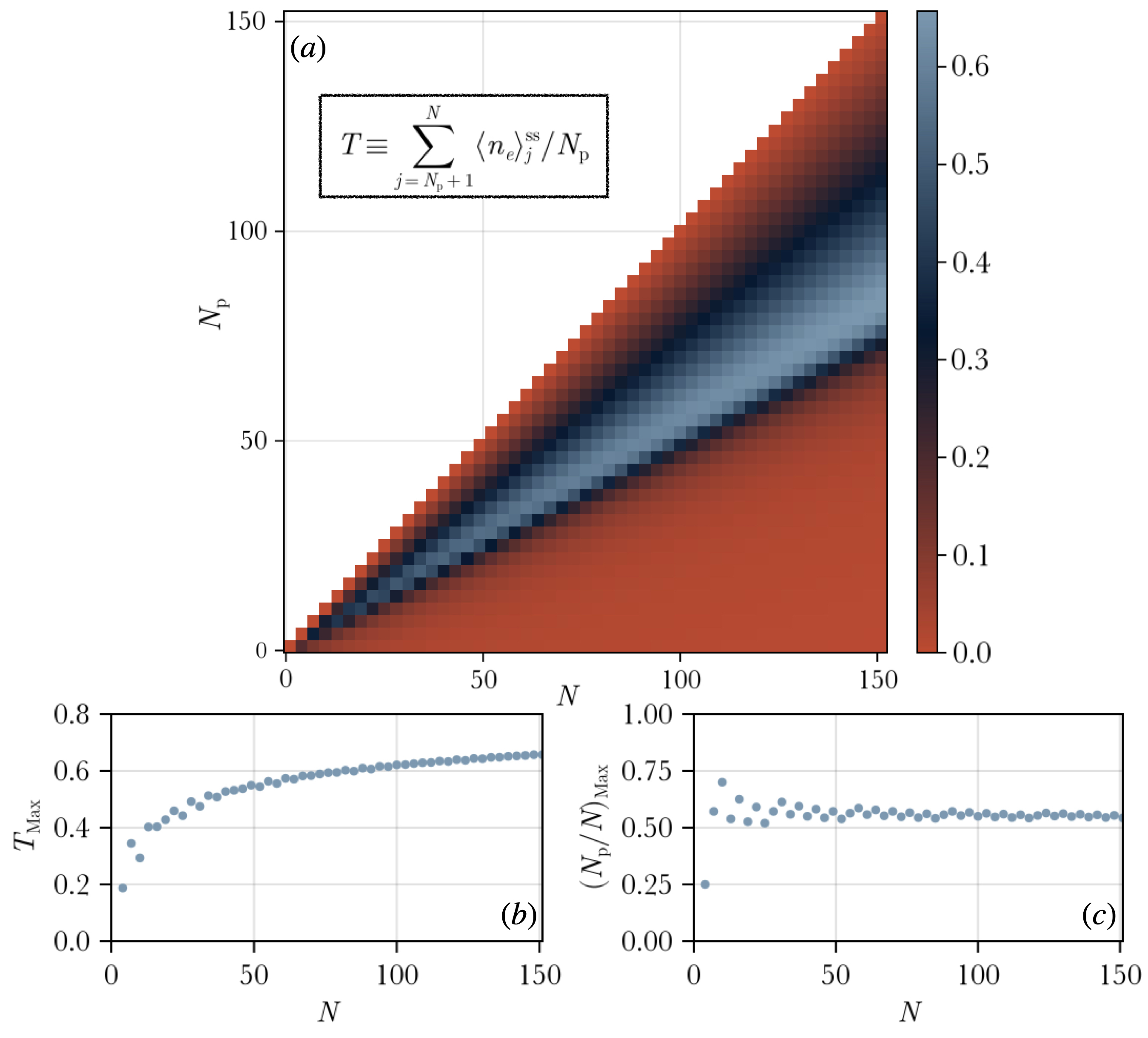}
    \caption{\textbf{Steady-state excitation transfer.} (a) Heatmap of the excitation transfer ratio $T$ as a function of total emitter number $N$ and number of initially pumped emitters $N_\text{p}$. A clear transition occurs near the critical point $N_\text{p} = N/2$, where the dark-states no longer localize significant excitations in the pumped ensemble, resulting in enhanced excitation transfer. (b) The maximal transfer ratio $T_\text{Max}$ as a function of total emitter number $N$. The peak value increases slowly with system size and shows no sign of saturation. (c) The ratio $(N_\text{p}/N)_\text{Max}$ at which maximal transfer occurs, plotted against $N$. This optimal pumped fraction converges toward approximately $0.55$ for large system sizes, slightly above the half-pumped threshold.}
    \label{energy_transfer}
\end{figure}

\section{Steady-state excitation transfer}
Figure~\ref{time_evolution} illustrates that excitations can be transferred from the initially pumped ensemble to the non-pumped one, where they are stored via $M$-excitation dark states. In this section, we quantitatively analyze the excitation transfer efficiency by defining the transfer ratio, $T \equiv \sum_{j=N_\text{p}+1}^N \langle n_e \rangle_j^\text{ss} / N_\text{p}$, which measures the fraction of initial excitations in the pumped ensemble that ultimately populate the un-pumped ensemble in the steady state.

\begin{figure*}[t]
    \centering
    \includegraphics[width=0.95\linewidth]{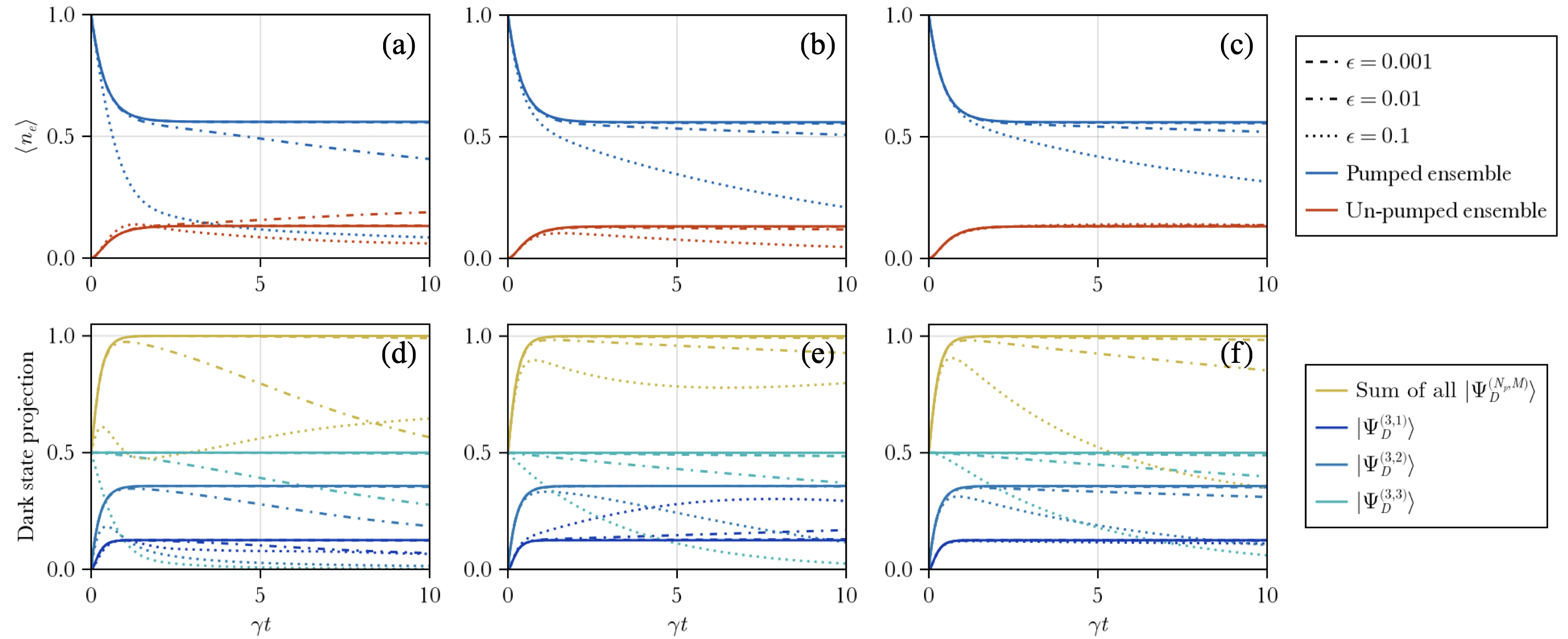}
    \caption{\textbf{Time evolution under system imperfections.} (a–c) Time evolution of the mean excitation number $\langle n_e \rangle$ in the pumped (blue) and un-pumped (red) ensembles for varying levels of (a) positional disorder, (b) nonradiative decay, and (c) dephasing, modeled respectively by $\epsilon = 0.001$ (dashed), $0.01$ (dash-dotted), and $0.1$ (dotted), where $\epsilon$ characterizes either the spatial deviation or the decay/dephasing rate relative to the collective decay rate $\gamma$. (d–f) Corresponding dark-state projections onto $|\psi_D^{(3,1)}\rangle$, $|\psi_D^{(3,2)}\rangle$, and $|\psi_D^{(3,3)}\rangle$ under each imperfection scenario. The yellow curves show the sum of projections onto all $|\psi_D^{(3,M)}\rangle$ dark states, including the zero-excitation ground state, thereby representing the total population within the ideal dark-state manifold (constructed according to Eq.~(\ref{DS_expression})). Notably, even when the total dark-state projection decreases, the excitation transfer remains effective—indicating that excitations are redistributed into other subradiant states emerging from symmetry breaking. The curves for positional disorder are averaged over 200 random positional disorder sampled from a Gaussian distribution.}
    \label{imperfection}
\end{figure*}

In Fig.~\ref{energy_transfer}(a), we plot $T$ as a function of the number of initially pumped emitters $N_\text{p}$ for various total number of emitter $N$. For $N_\text{p} < N/2$, the excitation remains largely localized in the pumped ensemble, resulting in a negligible transfer ratio ($T \approx 0$), consistent with excitation localization effects reported in previous works~\cite{poshakinskiy2021dimerization, fasser2024subradiance}. However, once the number of pumped emitters exceeds the critical threshold $N_\text{p} = N/2$, the system begins to transfer a significant portion of the excitations to the un-pumped ensemble.

This transition can be understood by considering the role of dark states: the number of available $M$-excitation dark states is limited by $M \leq \min\{N_\text{p}, N_\text{np}\}$. When $N_\text{p} > N/2$, the system contains more excitations than the dark states can store initially, and also the dark-states tend to localize excitations in the un-pumped ensemble in the steady-state. As a result, excess excitations either decay through collective radiation channels or are transferred to the un-pumped ensemble via the dissipative coupling. As $N_\text{p}$ increases further beyond $N/2$, fewer dark states are available, and the effect of collective decay becomes stronger. Eventually, when $N_\text{p} = N$, the un-pumped ensemble vanishes, and the system reduces to a fully symmetric Dicke configuration where all excitations decay via the superradiant channel, again leading to $T \rightarrow 0$.

The interplay between excitation delocalization (which promotes transfer) and the limited number of dark states (which preserve excitation) results in a peak in the transfer ratio for $N_\text{p} > N/2$. By extracting the maximal excitation transfer for each $N$, we observe in Fig. \ref{energy_transfer}(b) that the maximal transfer ratio $T_\text{Max}$ increases slowly with system size, without signs of saturation, indicating a persistent enhancement in excitation transport capacity as the system scales up. To further quantify the optimal pumping condition for maximal excitation transfer, we analyze the ratio $(N_\text{p}/N)_\text{Max}$ at which $T_\text{Max}$ occurs, as shown in Fig.~\ref{energy_transfer}(c). Interestingly, this optimal pumping fraction exhibits a consistent convergence behavior: while it fluctuates slightly for small system sizes due to the finite size effect, it stabilizes around $(N_\text{p}/N)_\text{Max} \approx 0.55$ as $N$ increases. This suggests that the most efficient excitation transfer in large systems occurs slightly above the critical half-pumped point, reflecting a subtle balance between dark-state saturation and collective dissipation pathways.


\section{Positional disorder, nonradiative decay, and dephasing effects}
In realistic systems, imperfections such as positional disorder, nonradiative decay, and dephasing inevitably affect the dynamics of excitation transfer. These imperfections open additional decay channels, causing excitations to leak out of the ideal dark-state manifold and potentially reducing storage efficiency. In this section, we analyze the system's robustness against such imperfections by simulating the dynamics for a representative configuration of $(N, N_\text{p}) = (8, 3)$, as shown in Fig.~\ref{imperfection}.

To model positional disorder, we assume the emitters deviate from their ideal positions according to a Gaussian distribution with standard deviation $\epsilon$ (in units of $\lambda_0$). As seen in Figs.~\ref{imperfection}(a) and \ref{imperfection}(d), the system remains largely unaffected for $\epsilon = 0.001$, a level of spatial precision achievable in superconducting qubit platforms. However, with increasing disorder (e.g., $\epsilon = 0.01$ and $0.1$), we observe a progressive degradation in dark-state population over time. Interestingly, for $\epsilon = 0.01$, while the projections onto individual dark states decay, the mean excitation number in the un-pumped ensemble increases compared to the disorder-free case. This counterintuitive effect arises because the additional decay channels introduced by disorder enable excitations to partially delocalize between ensembles, enhancing transfer before full dissipation occurs.

Nonradiative decay and dephasing—modeled respectively by rates $\gamma_\text{nr} = \epsilon\gamma$ and $\gamma_\phi = \epsilon\gamma$ for each emitter—introduce incoherent loss and dephasing effects. These are incorporated by modifying both the effective Hamiltonian and the master equation: $\hat{\mathcal{H}}_\text{eff} \rightarrow \hat{\mathcal{H}}_\text{eff} - i\gamma_\text{nr} + 2 \gamma_{\phi} \sum_n \hat{\sigma}_n^\dagger \hat{\sigma}_n / 2$, and $\dot{\hat{\rho}} \rightarrow \dot{\hat{\rho}} + \gamma_\text{nr}\sum_n \hat{\sigma}_n \hat{\rho}\hat{\sigma}_n^\dagger + 2\gamma_\phi \sum_n \hat{\sigma}_n^\dagger \hat{\sigma}_n \hat{\rho} \hat{\sigma}_n^\dagger \hat{\sigma}_n$ \cite{holzinger2022control}. Their effects are shown in Figs.~\ref{imperfection}(b,e) and \ref{imperfection}(c,f), respectively. Nonradiative decay primarily affects higher-excitation dark states, as seen in Fig.~\ref{imperfection}(e), where the population in $|\psi_D^{(3,3)}\rangle$ declines first. However, part of this population is redistributed into lower-excitation dark states, which remain populated longer. The repopulation of excitations from higher dark states to lower ones makes the overall transfer process robust to nonradiative loss and dephasing over finite timescales, in the sense that significant excitation transfer still occurs in the presence of system imperfections. A similar trend appears under dephasing [Figs.~\ref{imperfection}(c,f)], where the dark-state projections decay more gradually.

In Figs.~\ref{imperfection}(d–f), we also include a yellow curve representing the total projection onto all $|\psi_D^{(3,M)}\rangle$ dark states. While this total projection remains close to unity for weak imperfections, it decreases notably as $\epsilon$ increases. Importantly, although the mean excitation transfer remains nearly unchanged across different imperfections, this does not imply that the system's dynamics can still be described solely by the ideal dark-states. Instead, the stored excitations gradually shift into more general, long-lived subradiant states that emerge from symmetry-breaking imperfections. This resilience to loss and dephasing suggests that the excitation storage and transfer mechanism remains robust on current experimental platforms, such as quantum dots coupled with photonic crystal waveguides \cite{arcari2014near, sollner2015deterministic} or superconducting qubits coupled with microwave circuits \cite{sheremet2023waveguide, astafiev2010resonance}.

\section{Challenges of systems with lower symmetry}
Our analytical framework relies critically on the high degree of permutation symmetry in the system, particularly the condition that the coherent DDI terms in the effective Hamiltonian vanish entirely. This condition is satisfied, for example, in the so-called “mirror configuration,” where the emitter spacing is an integer multiple of the transition wavelength. In such cases, the system exhibits a single collective decay channel and uniform interaction strengths, allowing us to construct symmetrized $M$-excitation dark states and obtain compact analytic expressions for their time evolution.

\begin{figure}[t]
    \centering
    \includegraphics[width=1\linewidth]{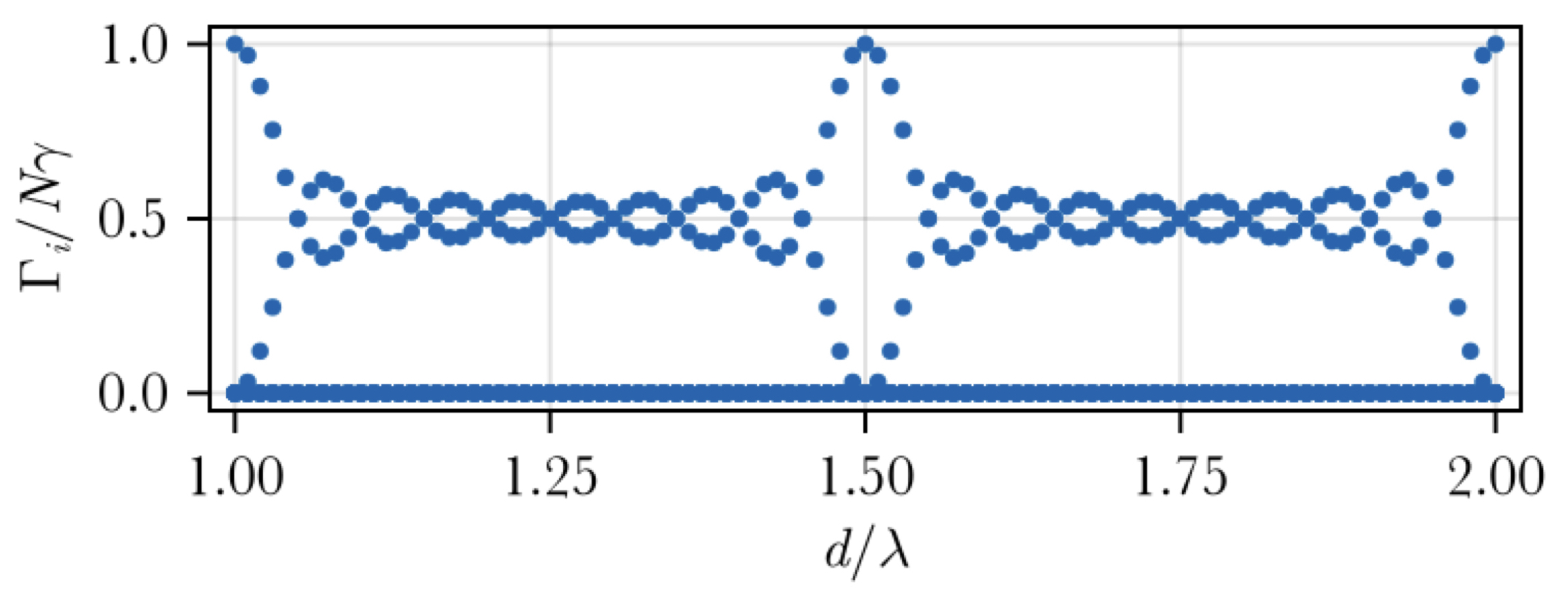}
    \caption{\textbf{Decay rates for different emitter spacings.} The dissipative interaction in Eq.~(\ref{master_eq}), where $\gamma_{m,n} = \gamma \cos(k_0 x_{m,n})$, can be diagonalized into a set of $N$ jump operators $\mathcal{J}_i$ with corresponding decay rates $\Gamma_i$, which are the eigenvalues of the dissipative interaction matrix $\gamma_{m,n}$~\cite{clemens2003collective, masson2022universality}. The figure shows the decay rates for $N=10$ under different emitter spacings $d$.}
    \label{decay_rates}
\end{figure}

However, in general WQED systems with non-vanishing coherent DDI, the situation becomes significantly more complex. The presence of coherent DDI introduces inhomogeneous energy shifts among emitters, breaking the permutation symmetry of the system. As a result, finding the zero-eigenstates of the effective Hamiltonian or the collective jump operators becomes a challenging task. Moreover, the presence of multiple decay channels further complicates the isolation and characterization of dark states within each excitation manifold. Fig.~\ref{decay_rates} illustrates the decay rates for different collective decay channels in a 1D WQED system with $N = 10$ emitters. The number of non-zero decay rates exceeds 1 when the emitter spacing deviates from $d = n\lambda/2$ (for $n = 1, 2, 3, \ldots$).

These challenges highlight important directions for future work. One promising path is to extend the current dark-state projection framework to systems in which the non-Hermitian part of the effective Hamiltonian dominates over the Hermitian part (i.e., systems with weak coherent DDI), allowing for the identification of extremely subradiant states (quasi-dark states) to investigate quasi-steady-state behaviors of the system.

\section{Conclusion}
We have explored excitation transfer dynamics in 1D waveguide QED systems mediated by collective dissipative interactions. By constructing a set of symmetrized $M$-excitation dark states for a highly symmetric emitter configuration, we derived analytical solutions for their time-evolution projections, which significantly reduce computational cost compared to full master equation simulations. While this approach relies on a high degree of permutation symmetry and is not directly extensible to arbitrary geometries, it offers an exact and powerful solution for an important class of analytically tractable WQED systems. These dark states provide a compact representation of the system's dynamics and capture essential features of excitation transport between two spatially separated emitter ensembles.

We identified the optimal pumping condition that maximizes this transfer, and found that for large system sizes, the optimal pumping fraction $(N_\text{p}/N)_\text{Max}$ converges to approximately 0.55—slightly above the intuitive half-pumping threshold. This behavior reflects the competition between the number of accessible dark states and collective decay channels. Furthermore, we examined the robustness of excitation transfer against realistic system imperfections, including positional disorder, nonradiative decay, and dephasing. Our analysis shows that while high levels of imperfection can disrupt dark-state populations, moderate levels—achievable in state-of-the-art experiments—preserve the efficiency of excitation transfer and maintain strong projection onto lower-excitation dark states over experimentally relevant timescales.

Our findings highlight the utility of many-body dark states as a resource for excitation transport and suggest that robust quantum control is possible in dissipatively coupled emitter systems. These insights could aid the design of quantum devices for storage, transfer, and processing of quantum information in integrated WQED platforms.

\section{Acknowledgments}
We acknowledge support from the National Science and Technology Council (NSTC), Taiwan, under the Grant No. NSTC-112-2119-M-001-007 and Grant No. NSTC-112-2112-M-001-079-MY3, and from Academia Sinica under Grant AS-CDA-113-M04. We are also grateful for support from TG 1.2 of NCTS at NTU. GDL acknowledges support from Grant No. NSTC-113-2112-M-002 -025 and No. NSTC-112-2112-M-002 -001.

\begin{widetext}
\appendix*
\section{Derivation of the $M$-excitation dark state}
Considering the mirror configuration with $d = n\lambda_0$, where $n$ is an integer and $\lambda_0$ is the resonant wavelength, the effective Hamiltonian is given by $\hat{\mathcal{H}}_\text{eff} = -\frac{i\gamma}{2}\mathcal{S}^\dagger\mathcal{S}$. By grouping the emitters into pumped and un-pumped ensembles, containing $N_\text{p}$ and $N - N_\text{p}$ emitters respectively, the symmetrized zero-energy eigenstates of the system with $M$ excitations, satisfying $\hat{\mathcal{H}}_\text{eff} \ket{\Psi_D^{(N_\text{p}, M)}} = 0$, are also zero-energy eigenstates of the collective lowering operator $\mathcal{S} = \mathcal{S}_\text{p} + \mathcal{S}_\text{np}$:
\begin{equation}
     \mathcal{S} \ket{\Psi_D^{(N_p, M)}} = 0\ket{\Psi_D^{(N_p, M)}}.
\label{eigenvalue_eq}
\end{equation}
To construct the $M$-excitation dark state, we adopt the following ansatz \cite{holzinger2022control}:
\begin{equation}
    \ket{\Psi_D^{(N_p, M)}} = \sum_{k=0}^M (-1)^k r_k\left( \mathcal{S}^\dagger_{p}\right)^{M-k}\left( \mathcal{S}^\dagger_{np}\right)^{k} \ket{G}.
\end{equation}
Now Eq.~(\ref{eigenvalue_eq}) becomes $M+1$ coupled equations and the coefficients are solvable:
\begin{align}
    \mathcal{S} \ket{\Psi_D^{(N_p, M)}} =& \left[ \sum_{k=0}^M(-1)^k r_k\left(\mathcal{S}_\text{p}\mathcal{S}^\dagger_\text{p} \right) \left( \mathcal{S}^\dagger_\text{p} \right)^{M-k-1} \left( \mathcal{S}^\dagger_\text{np} \right)^k
    +
    \sum_{k=0}^M(-1)^k r_k\left(\mathcal{S}_\text{np}\mathcal{S}^\dagger_\text{np} \right) \left( \mathcal{S}^\dagger_\text{p} \right)^{M-k} \left( \mathcal{S}^\dagger_\text{np} \right)^{k-1} \right] \ket{G}, \notag \\
    =& 0.
\end{align}
By matching the terms with the same excitation number, we obtain:
\begin{align}
    &\left[ (-1)^{k-1} r_{k-1} \left( \mathcal{S}_\text{p}\mathcal{S}^\dagger_\text{p} \right) + (-1)^k r_k \left( \mathcal{S}_\text{np}\mathcal{S}^\dagger_\text{np} \right) \right] \left( \mathcal{S}^\dagger_\text{p} \right)^{M-k} \left(\mathcal{S}^\dagger_\text{np} \right)^{k-1} \ket{G}, \notag\\
    \propto &  
    \left[ r_{k-1} \left( \mathcal{S}_\text{p}\mathcal{S}^\dagger_\text{p} \right) - r_k \left( \mathcal{S}_\text{np}\mathcal{S}^\dagger_\text{np} \right) \right] \ket{N_\text{p}/2, m_{\text{p}, M-k}} \otimes \ket{N_\text{np}/2, m_{\text{np}, k-1}} = 0,
\end{align}
where $\ket{N_\text{p}/2, m_{\text{p}, M-k}} \otimes \ket{N_\text{np}/2, m_{\text{np}, k-1}}$ denotes Dicke states for the pumped and unpumped ensembles with $M-k$ and $k-1$ excitations, respectively. Here, $m_{\text{p}, M-k} = -N_\text{p}/2 + (M-k)$ and $m_{\text{np}, k-1} = -N_\text{np}/2 + (k-1)$. Using the action of the collective operators on Dicke states \cite{gross1982superradiance}:
\begin{align}
    \mathcal{S}^\dagger\ket{N/2, m} &= A_m \ket{N/2, m+1}, \notag \\
    \mathcal{S}\ket{N/2, m} &= A_{m-1}\ket{N/2, m-1},
\end{align}
where $A_m = \sqrt{(N/2)(N/2+1) - m(m+1)}$, we can derive the recursion relation for the coefficient, $r_k$:
\begin{equation}
    r_{k-1} A^2_{-N_\text{p}/2+(M-k)} - r_kA^2_{-(N-N_\text{p})/2+(k-1)} = 0 \longrightarrow r_k = \frac{(N_\text{p} - M+k)(M-k+1)}{k(N-N_\text{p}-k+1)}r_{k-1}.
    \label{recursion}
\end{equation}
The $M$-excitation dark states can then be constructed up to normalization.
Another derivation is by symmetrizing the $M$-excitation antisymmetric states:
\begin{equation}
    \ket{\Psi_\text{D}} \propto (\sigma_i^\dagger - \sigma_j^\dagger)...(\sigma_m^\dagger - \sigma_n^\dagger)\ket{G}.
\end{equation}
Taking the $M=2$-excitation dark state as an example, we have:
\begin{align}
    \ket{\Psi^{(N_\text{p}, 2)}_\text{D}} &\propto \sum_{(i, \alpha) = (1, 1)}^{(N_\text{p}, N_\text{np})} \sum_{\substack{(j, \beta) = (1, 1) \\ (j, \beta) \ne (i, \alpha)}}^{(N_\text{p}, N_\text{np})}
    \left(\sigma^\dagger_i - \sigma^\dagger_\alpha \right) \left(\sigma^\dagger_j - \sigma^\dagger_\beta \right)\ket{G}, \notag \\ &= 
    \left[
    \begin{pmatrix}
    N_\text{np} \\
    2
    \end{pmatrix}
    \left( \sum_{\substack{i,j=1\\i\ne j}}^{N_\text{p}} \sigma_i^\dagger \sigma_j^\dagger \right) -
    \begin{pmatrix}
    N_\text{np} -1\\
    1
    \end{pmatrix}
    \begin{pmatrix}
    N_\text{p} -1\\
    1
    \end{pmatrix}
    \left( \sum_{i = 1}^{N_\text{p}} \sigma_i^\dagger \right)\left( \sum_{\alpha = 1}^{N_\text{np}} \sigma_\alpha^\dagger \right)
    +
    \begin{pmatrix}
    N_\text{p} \\
    2
    \end{pmatrix}\left( \sum_{\substack{\alpha,\beta=1\\ \alpha \ne  \beta}}^{N_\text{np}}\sigma_\alpha^\dagger \sigma_\beta^\dagger \right) \right] \ket{G},\notag\\ 
    &= 
    \left[
    \begin{pmatrix}
    N_\text{np} \\
    2
    \end{pmatrix}
    \left( \mathcal{S}_\text{p}^\dagger \right)^2 -
    \begin{pmatrix}
    N_\text{np} -1\\
    1
    \end{pmatrix}
    \begin{pmatrix}
    N_\text{p} -1\\
    1
    \end{pmatrix}
    \mathcal{S}^\dagger_\text{p} \mathcal{S}^\dagger_\text{np}
    +
    \begin{pmatrix}
    N_\text{p} \\
    2
    \end{pmatrix}\left( \mathcal{S}_\text{np}^\dagger \right)^2 \right] \ket{G}.
\end{align}
The binomial coefficients represent the number of ways to distribute excitations between ensembles for each configuration. The general form of the $M$-excitation dark state is given in Eq.~(\ref{DS_expression}), and can be verified to satisfy the recursion relation in Eq.~(\ref{recursion}):
\begin{align}
    &\begin{pmatrix}
    N_\text{p} - M +k -1\\
    k-1
    \end{pmatrix}
    \begin{pmatrix}
    N-N_\text{p}-k+1\\
    M-k+1
    \end{pmatrix}  
    A^2_{-N_\text{p}/2+(M-k)} - 
    \begin{pmatrix}
    N_\text{p}-M+k\\
    k
    \end{pmatrix} 
    \begin{pmatrix}
    N-N_\text{p}-k\\
    M-k
    \end{pmatrix}
    A^2_{-N_\text{np}/2+(k-1)}, \notag \\
    \propto &
    \left( \frac{N-N_\text{p}-k+1}{M-k+1}A^2_{-N_\text{p}/2+(M-k)}\right) - \left(\frac{N_\text{p}-M+k}{k}A^2_{-N_\text{np}/2+(k-1)} \right), \notag \\
    =&
    \left[ \frac{N-N_\text{p}-k+1}{M-k+1}(N_\text{p}-M+k)(M-k+1) \right] - \left[ \frac{N_\text{p}-M+k}{k}(N-N_\text{p}-k+1)k \right] = 0.
\end{align}

\end{widetext}

\bibliographystyle{unsrt}
\bibliography{ref}

\begin{thebibliography}{10}

\bibitem{sheremet2023waveguide}
Alexandra~S Sheremet, Mihail~I Petrov, Ivan~V Iorsh, Alexander~V Poshakinskiy, and Alexander~N Poddubny.
\newblock Waveguide quantum electrodynamics: Collective radiance and photon-photon correlations.
\newblock {\em Reviews of Modern Physics}, 95(1):015002, 2023.

\bibitem{lodahl2017chiral}
Peter Lodahl, Sahand Mahmoodian, S{\o}ren Stobbe, Arno Rauschenbeutel, Philipp Schneeweiss, J{\"u}rgen Volz, Hannes Pichler, and Peter Zoller.
\newblock Chiral quantum optics.
\newblock {\em Nature}, 541(7638):473--480, 2017.

\bibitem{arcari2014near}
Marta Arcari, Immo S{\"o}llner, Alisa Javadi, S~Lindskov~Hansen, Sahand Mahmoodian, Jin Liu, Henri Thyrrestrup, Eun~Hye Lee, Jin~Dong Song, S{\o}ren Stobbe, et~al.
\newblock Near-unity coupling efficiency of a quantum emitter to a photonic crystal waveguide.
\newblock {\em Physical Review Letters}, 113(9):093603, 2014.

\bibitem{le2021experimental}
Hanna Le~Jeannic, Tom{\'a}s Ramos, Signe~F Simonsen, Tommaso Pregnolato, Zhe Liu, R{\"u}diger Schott, Andreas~D Wieck, Arne Ludwig, Nir Rotenberg, Juan~Jos{\'e} Garc{\'\i}a-Ripoll, et~al.
\newblock Experimental reconstruction of the few-photon nonlinear scattering matrix from a single quantum dot in a nanophotonic waveguide.
\newblock {\em Physical Review Letters}, 126(2):023603, 2021.

\bibitem{douglas2015quantum}
James~S Douglas, Hessam Habibian, C-L Hung, Alexey~V Gorshkov, H~Jeff Kimble, and Darrick~E Chang.
\newblock Quantum many-body models with cold atoms coupled to photonic crystals.
\newblock {\em Nature Photonics}, 9(5):326--331, 2015.

\bibitem{chang2018colloquium}
DE~Chang, JS~Douglas, Alejandro Gonz{\'a}lez-Tudela, C-L Hung, and HJ~Kimble.
\newblock Colloquium: Quantum matter built from nanoscopic lattices of atoms and photons.
\newblock {\em Reviews of Modern Physics}, 90(3):031002, 2018.

\bibitem{mahmoodian2020dynamics}
Sahand Mahmoodian, Giuseppe Calaj{\'o}, Darrick~E Chang, Klemens Hammerer, and Anders~S S{\o}rensen.
\newblock Dynamics of many-body photon bound states in chiral waveguide qed.
\newblock {\em Physical Review X}, 10(3):031011, 2020.

\bibitem{iversen2022self}
Ole~A Iversen and Thomas Pohl.
\newblock Self-ordering of individual photons in waveguide qed and rydberg-atom arrays.
\newblock {\em Physical Review Research}, 4(2):023002, 2022.

\bibitem{gonzalez2024light}
Alejandro Gonz{\'a}lez-Tudela, Andreas Reiserer, Juan~Jos{\'e} Garc{\'\i}a-Ripoll, and Francisco~J Garc{\'\i}a-Vidal.
\newblock Light--matter interactions in quantum nanophotonic devices.
\newblock {\em Nature Reviews Physics}, 6(3):166--179, 2024.

\bibitem{le2005nanofiber}
Fam Le~Kien, S~Dutta Gupta, KP~Nayak, and K~Hakuta.
\newblock Nanofiber-mediated radiative transfer between two distant atoms.
\newblock {\em Physical Review A—Atomic, Molecular, and Optical Physics}, 72(6):063815, 2005.

\bibitem{kien2008cooperative}
Fam~Le Kien and K~Hakuta.
\newblock Cooperative enhancement of channeling of emission from atoms into a nanofiber.
\newblock {\em Physical Review A—Atomic, Molecular, and Optical Physics}, 77(1):013801, 2008.

\bibitem{van2013photon}
Arjan~F Van~Loo, Arkady Fedorov, Kevin Lalumiere, Barry~C Sanders, Alexandre Blais, and Andreas Wallraff.
\newblock Photon-mediated interactions between distant artificial atoms.
\newblock {\em Science}, 342(6165):1494--1496, 2013.

\bibitem{goban2015superradiance}
A~Goban, C-L Hung, JD~Hood, S-P Yu, JA~Muniz, O~Painter, and HJ~Kimble.
\newblock Superradiance for atoms trapped along a photonic crystal waveguide.
\newblock {\em Physical Review Letters}, 115(6):063601, 2015.

\bibitem{pichler2015quantum}
Hannes Pichler, Tom{\'a}s Ramos, Andrew~J Daley, and Peter Zoller.
\newblock Quantum optics of chiral spin networks.
\newblock {\em Physical Review A}, 91(4):042116, 2015.

\bibitem{shahmoon2016highly}
Ephraim Shahmoon, Pjotrs Gri{\v{s}}ins, Hans~Peter Stimming, Igor Mazets, and Gershon Kurizki.
\newblock Highly nonlocal optical nonlinearities in atoms trapped near a waveguide.
\newblock {\em Optica}, 3(7):725--733, 2016.

\bibitem{ruostekoski2016emergence}
Janne Ruostekoski and Juha Javanainen.
\newblock Emergence of correlated optics in one-dimensional waveguides for classical and quantum atomic gases.
\newblock {\em Physical Review Letters}, 117(14):143602, 2016.

\bibitem{ruostekoski2017arrays}
Janne Ruostekoski and Juha Javanainen.
\newblock Arrays of strongly coupled atoms in a one-dimensional waveguide.
\newblock {\em Physical Review A}, 96(3):033857, 2017.

\bibitem{holzinger2022control}
Raphael Holzinger, Ricardo Gutierrez-Jauregui, Teresa H{\"o}nigl-Decrinis, Gerhard Kirchmair, Ana Asenjo-Garcia, and Helmut Ritsch.
\newblock Control of localized single-and many-body dark states in waveguide qed.
\newblock {\em Physical Review Letters}, 129(25):253601, 2022.

\bibitem{asenjo2017exponential}
Ana Asenjo-Garcia, M~Moreno-Cardoner, Andreas Albrecht, HJ~Kimble, and Darrick~E Chang.
\newblock Exponential improvement in photon storage fidelities using subradiance and “selective radiance” in atomic arrays.
\newblock {\em Physical Review X}, 7(3):031024, 2017.

\bibitem{plankensteiner2015selective}
David Plankensteiner, Laurin Ostermann, Helmut Ritsch, and Claudiu Genes.
\newblock Selective protected state preparation of coupled dissipative quantum emitters.
\newblock {\em Scientific reports}, 5(1):16231, 2015.

\bibitem{moreno2019subradiance}
Maria Moreno-Cardoner, David Plankensteiner, Laurin Ostermann, Darrick~E Chang, and Helmut Ritsch.
\newblock Subradiance-enhanced excitation transfer between dipole-coupled nanorings of quantum emitters.
\newblock {\em Physical Review A}, 100(2):023806, 2019.

\bibitem{masson2020atomic}
Stuart~J Masson and Ana Asenjo-Garcia.
\newblock Atomic-waveguide quantum electrodynamics.
\newblock {\em Physical Review Research}, 2(4):043213, 2020.

\bibitem{ferioli2021storage}
Giovanni Ferioli, Antoine Glicenstein, Loic Henriet, Igor Ferrier-Barbut, and Antoine Browaeys.
\newblock Storage and release of subradiant excitations in a dense atomic cloud.
\newblock {\em Physical Review X}, 11(2):021031, 2021.

\bibitem{needham2019subradiance}
Jemma~A Needham, Igor Lesanovsky, and Beatriz Olmos.
\newblock Subradiance-protected excitation transport.
\newblock {\em New Journal of Physics}, 21(7):073061, 2019.

\bibitem{zanner2022coherent}
Maximilian Zanner, Tuure Orell, Christian~MF Schneider, Romain Albert, Stefan Oleschko, Mathieu~L Juan, Matti Silveri, and Gerhard Kirchmair.
\newblock Coherent control of a multi-qubit dark state in waveguide quantum electrodynamics.
\newblock {\em Nature Physics}, 18(5):538--543, 2022.

\bibitem{lvovsky2009optical}
Alexander~I Lvovsky, Barry~C Sanders, and Wolfgang Tittel.
\newblock Optical quantum memory.
\newblock {\em Nature photonics}, 3(12):706--714, 2009.

\bibitem{zhao2009long}
R~Zhao, YO~Dudin, SD~Jenkins, CJ~Campbell, DN~Matsukevich, TAB Kennedy, and A~Kuzmich.
\newblock Long-lived quantum memory.
\newblock {\em Nature Physics}, 5(2):100--104, 2009.

\bibitem{zhao2009millisecond}
Bo~Zhao, Yu-Ao Chen, Xiao-Hui Bao, Thorsten Strassel, Chih-Sung Chuu, Xian-Min Jin, J{\"o}rg Schmiedmayer, Zhen-Sheng Yuan, Shuai Chen, and Jian-Wei Pan.
\newblock A millisecond quantum memory for scalable quantum networks.
\newblock {\em Nature Physics}, 5(2):95--99, 2009.

\bibitem{dinc2020multidimensional}
Fatih Dinc, Lauren~E Hayward, and Agata~M Bra{\'n}czyk.
\newblock Multidimensional super-and subradiance in waveguide quantum electrodynamics.
\newblock {\em Physical Review Research}, 2(4):043149, 2020.

\bibitem{almanakly2022towards}
Aziza Almanakly.
\newblock {\em Towards a Quantum Network with Waveguide Quantum Electrodynamics}.
\newblock PhD thesis, Massachusetts Institute of Technology, 2022.

\bibitem{stannigel2012driven}
Kai Stannigel, Peter Rabl, and Peter Zoller.
\newblock Driven-dissipative preparation of entangled states in cascaded quantum-optical networks.
\newblock {\em New Journal of Physics}, 14(6):063014, 2012.

\bibitem{ramos2014quantum}
Tom{\'a}s Ramos, Hannes Pichler, Andrew~J Daley, and Peter Zoller.
\newblock Quantum spin dimers from chiral dissipation in cold-atom chains.
\newblock {\em Physical Review Letters}, 113(23):237203, 2014.

\bibitem{quach2022superabsorption}
James~Q Quach, Kirsty~E McGhee, Lucia Ganzer, Dominic~M Rouse, Brendon~W Lovett, Erik~M Gauger, Jonathan Keeling, Giulio Cerullo, David~G Lidzey, and Tersilla Virgili.
\newblock Superabsorption in an organic microcavity: Toward a quantum battery.
\newblock {\em Science advances}, 8(2):eabk3160, 2022.

\bibitem{ferraro2018high}
Dario Ferraro, Michele Campisi, Gian~Marcello Andolina, Vittorio Pellegrini, and Marco Polini.
\newblock High-power collective charging of a solid-state quantum battery.
\newblock {\em Physical Review Letters}, 120(11):117702, 2018.

\bibitem{quach2020using}
James~Q Quach and William~J Munro.
\newblock Using dark states to charge and stabilize open quantum batteries.
\newblock {\em Physical Review Applied}, 14(2):024092, 2020.

\bibitem{dicke1954coherence}
Robert~H Dicke.
\newblock Coherence in spontaneous radiation processes.
\newblock {\em Physical Review}, 93(1):99, 1954.

\bibitem{gross1982superradiance}
Michel Gross and Serge Haroche.
\newblock Superradiance: An essay on the theory of collective spontaneous emission.
\newblock {\em Physics reports}, 93(5):301--396, 1982.

\bibitem{albrecht2019subradiant}
Andreas Albrecht, Lo{\"\i}c Henriet, Ana Asenjo-Garcia, Paul~B Dieterle, Oskar Painter, and Darrick~E Chang.
\newblock Subradiant states of quantum bits coupled to a one-dimensional waveguide.
\newblock {\em New Journal of Physics}, 21(2):025003, 2019.

\bibitem{bettles2016cooperative}
Robert~J Bettles, Simon~A Gardiner, and Charles~S Adams.
\newblock Cooperative eigenmodes and scattering in one-dimensional atomic arrays.
\newblock {\em Physical Review A}, 94(4):043844, 2016.

\bibitem{ballantine2021quantum}
KE~Ballantine and J~Ruostekoski.
\newblock Quantum single-photon control, storage, and entanglement generation with planar atomic arrays.
\newblock {\em PRX Quantum}, 2(4):040362, 2021.

\bibitem{poshakinskiy2021dimerization}
Alexander~V Poshakinskiy and Alexander~N Poddubny.
\newblock Dimerization of many-body subradiant states in waveguide quantum electrodynamics.
\newblock {\em Physical Review Letters}, 127(17):173601, 2021.

\bibitem{fasser2024subradiance}
Martin Fasser, Laurin Ostermann, Helmut Ritsch, and Christoph Hotter.
\newblock Subradiance and superradiant long range excitation transport among quantum emitter ensembles in a waveguide.
\newblock {\em arXiv preprint arXiv:2405.07833}, 2024.

\bibitem{higgins2014superabsorption}
KDB Higgins, SC~Benjamin, TM~Stace, GJ~Milburn, Brendon~William Lovett, and EM~Gauger.
\newblock Superabsorption of light via quantum engineering.
\newblock {\em Nature communications}, 5(1):4705, 2014.

\bibitem{yang2021realization}
Daeho Yang, Seung-hoon Oh, Junseok Han, Gibeom Son, Jinuk Kim, Junki Kim, Moonjoo Lee, and Kyungwon An.
\newblock Realization of superabsorption by time reversal of superradiance.
\newblock {\em Nature Photonics}, 15(4):272--276, 2021.

\bibitem{okaba2019superradiance}
Shoichi Okaba, Deshui Yu, Luca Vincetti, Fetah Benabid, and Hidetoshi Katori.
\newblock Superradiance from lattice-confined atoms inside hollow core fibre.
\newblock {\em Communications Physics}, 2(1):136, 2019.

\bibitem{corzo2019waveguide}
Neil~V Corzo, J{\'e}r{\'e}my Raskop, Aveek Chandra, Alexandra~S Sheremet, Baptiste Gouraud, and Julien Laurat.
\newblock Waveguide-coupled single collective excitation of atomic arrays.
\newblock {\em Nature}, 566(7744):359--362, 2019.

\bibitem{mirhosseini2019cavity}
Mohammad Mirhosseini, Eunjong Kim, Xueyue Zhang, Alp Sipahigil, Paul~B Dieterle, Andrew~J Keller, Ana Asenjo-Garcia, Darrick~E Chang, and Oskar Painter.
\newblock Cavity quantum electrodynamics with atom-like mirrors.
\newblock {\em Nature}, 569(7758):692--697, 2019.

\bibitem{astafiev2010resonance}
O~Astafiev, Alexandre~M Zagoskin, AA~Abdumalikov~Jr, Yu~A Pashkin, T~Yamamoto, K~Inomata, Y~Nakamura, and Jaw~Shen Tsai.
\newblock Resonance fluorescence of a single artificial atom.
\newblock {\em Science}, 327(5967):840--843, 2010.

\bibitem{honigl2020two}
T~H{\"o}nigl-Decrinis, R~Shaikhaidarov, SE~De~Graaf, VN~Antonov, and OV~Astafiev.
\newblock Two-level system as a quantum sensor for absolute calibration of power.
\newblock {\em Physical Review Applied}, 13(2):024066, 2020.

\bibitem{solano2017super}
Pablo Solano, Pablo Barberis-Blostein, Fredrik~K Fatemi, Luis~A Orozco, and Steven~L Rolston.
\newblock Super-radiance reveals infinite-range dipole interactions through a nanofiber.
\newblock {\em Nature communications}, 8(1):1857, 2017.

\bibitem{lalumiere2013input}
Kevin Lalumiere, Barry~C Sanders, Arjan~F van Loo, Arkady Fedorov, Andreas Wallraff, and Alexandre Blais.
\newblock Input-output theory for waveguide qed with an ensemble of inhomogeneous atoms.
\newblock {\em Physical Review A—Atomic, Molecular, and Optical Physics}, 88(4):043806, 2013.

\bibitem{chang2012cavity}
Darrick~E Chang, L~Jiang, AV~Gorshkov, and HJ~Kimble.
\newblock Cavity qed with atomic mirrors.
\newblock {\em New Journal of Physics}, 14(6):063003, 2012.

\bibitem{sollner2015deterministic}
Immo S{\"o}llner, Sahand Mahmoodian, Sofie~Lindskov Hansen, Leonardo Midolo, Alisa Javadi, Gabija Kir{\v{s}}ansk{\.e}, Tommaso Pregnolato, Haitham El-Ella, Eun~Hye Lee, Jin~Dong Song, et~al.
\newblock Deterministic photon--emitter coupling in chiral photonic circuits.
\newblock {\em Nature nanotechnology}, 10(9):775--778, 2015.

\bibitem{clemens2003collective}
JP~Clemens, L~Horvath, BC~Sanders, and HJ~Carmichael.
\newblock Collective spontaneous emission from a line of atoms.
\newblock {\em Physical Review A}, 68(2):023809, 2003.

\bibitem{masson2022universality}
Stuart~J Masson and Ana Asenjo-Garcia.
\newblock Universality of dicke superradiance in arrays of quantum emitters.
\newblock {\em Nature Communications}, 13(1):2285, 2022.

\end{thebibliography}

\end{document}